# Room Temperature Anisotropic Photoresponse in Low-Symmetry van der Waals Semiconductor $CrPS_4$

Cédric A. Cordero-Silis,[1] Daniel Vaquero,[1] Teresa López-Carrasco,[1]
Harshan Madeshwaran,[1] and Marcos H. D. Guimarães[1,*]

[1] *Zernike Institute for Advanced Materials, University of Groningen, 9747 AG Groningen, The Netherlands*

The crystalline and optical anisotropy of low-symmetry two-dimensional (2D) materials can enable strong dichroic responses, enhancing polarization contrast for photonic and optoelectronic devices. Here, we unveil pronounced optical and optoelectronic anisotropy in chromium thiophosphate ($CrPS_4$) arising from the strong coupling between light polarization and its intrinsic crystal symmetry. Linearly polarized reflectivity and scanning photocurrent measurements in the 1.37–2.48 eV range reveal a robust dichroic response. The linear dichroism in reflection (RLD) reaches ∼50%, while in photocurrent (PCLD) it increases to ∼60%, with a sign reversal of the RLD between 1.6-1.8 eV, enabling strong narrow-band polarization contrast at room temperature. We attribute these anisotropic responses to the interaction between polarized light and $Cr^{3+}$ *d*-orbital $T_1$ and $T_2$ transitions. Spatially resolved photocurrent mapping further shows that the photocurrent is strongly dependent on the crystallographic axis: a 3-fold enhancement is obtained along the *b*-axis compared to the *a*-axis, yielding a clear 180° modulation of photoresponse across different contact orientations. Together, our findings establish $CrPS_4$ as a highly anisotropic 2D semiconductor with strong linear dichroism and polarization-sensitive photoresponse at room temperature. These characteristics highlight $CrPS_4$ as a promising platform for narrow-band polarized photodetectors, anisotropic photo-transport, and future 2D spintronic and magneto-optical devices.

**Corresponding author:**
[*] m.h.guimaraes@rug.nl

## I. INTRODUCTION

Recently, transition metal dichalcogenides (TMDs), notable for their strong absorption and direct band gap at the monolayer limit,[1,2] have been employed to enhance two-dimensional (2D) photodetector performance.[3–5] However, their most common crystal structure, the hexagonal 2H phase, does not present optical anisotropy in the linear response. In contrast, low-symmetry 2D van der Waals (vdWs) semiconductors have emerged as a unique platform where the spin degree-of-freedom, crystal anisotropy and light-matter interaction are intertwined. This grants new possibilities for controlling the optical response of photonic devices by exploiting their optoelectronic properties.[6–8] Therefore, the emergence of low-symmetry materials with anisotropic properties offers a novel pathway for compact and multifunctional polarization-sensitive optoelectronic devices without any additional optical component.[9–12]

Polarization-sensitive 2D photodetectors with dichroic photoresponse have been engineered using various low-symmetry materials. In the literature, the difference between the photoresponse for orthogonal light polarizations, referred to as photocurrent linear dichroism (PCLD), ranges values between ∼10%-60%, with the highest values reported at temperatures below 10 K.[13–15] Interestingly, a 90° phase shift in the PCLD polarization resolved photoresponse has been reported for different excitonic features in $ReS_2$, but only resolvable at cryogenic temperatures.[16] The strongest dichroic response in a vdWs material was reported for hBN encapsulated CrSBr (with *Pmmn* space group), where its quasi-1D nature boosts the PCLD up to ∼86% at low-temperatures.[17]

Chromium thiophosphate ($CrPS_4$) is a vdWs material with monoclinic symmetry, lower than the previously studied materials. Figure 1a shows a representative illustration of the $CrPS_4$ crystal along the *a*-*b* and *a*-*c* planes.[18,19] Earlier experimental reports and theoretical calculations indicate that the material belongs to the $C_2/m$ space group, nevertheless, recent X-ray measurements point towards a lower $C_2$ symmetry.[20–22] Recently, a lot of interest on this material has emerged due to its magnetic interactions, strong magnetoconductance modulation,[20,23–25] long-distance magnon transport,[26] and strain-dependent band structure.[27] Nonetheless, the low symmetry of the material leads to pronounced anisotropic optical properties, highlighting its importance for on-chip polarized photodetectors and proximitized devices.[28–31] More recent experiments have mainly focused on low-temperature photocurrent responses and their relation to magnetic phases.[32,33] However, the polarization dependence of the reflectivity and the photoresponse along different crystallographic directions remains underexplored, particularly at room temperature.

To address this, we perform room temperature reflection linear dichroism (RLD) and PCLD measurements in $CrPS_4$ devices with thicknesses between 11-25 nm. We observe a strong anisotropic photoresponse for laser energies in the region between 1.6 eV and 1.9 eV, with a strong dichroic response – up to 50% polarization contrast in reflection and 60% in photocurrent. These values are comparable with other 2D-based linearly polarized photodetectors but with the advantage of room temperature operation over cryogenic temperatures.[13–17] In addition to that, we perform scanning photocurrent measurements along different crystallographic directions of the material. In this configuration, we measure the total generated photocurrent and observe a 3-fold enhancement of the intensity along the b-axis with respect to the photoresponse along the a-axis. Our measurements demonstrate the strong resistance anisotropy in this material showcasing the role of the crystal symmetry in the photogenerated carriers across the device.

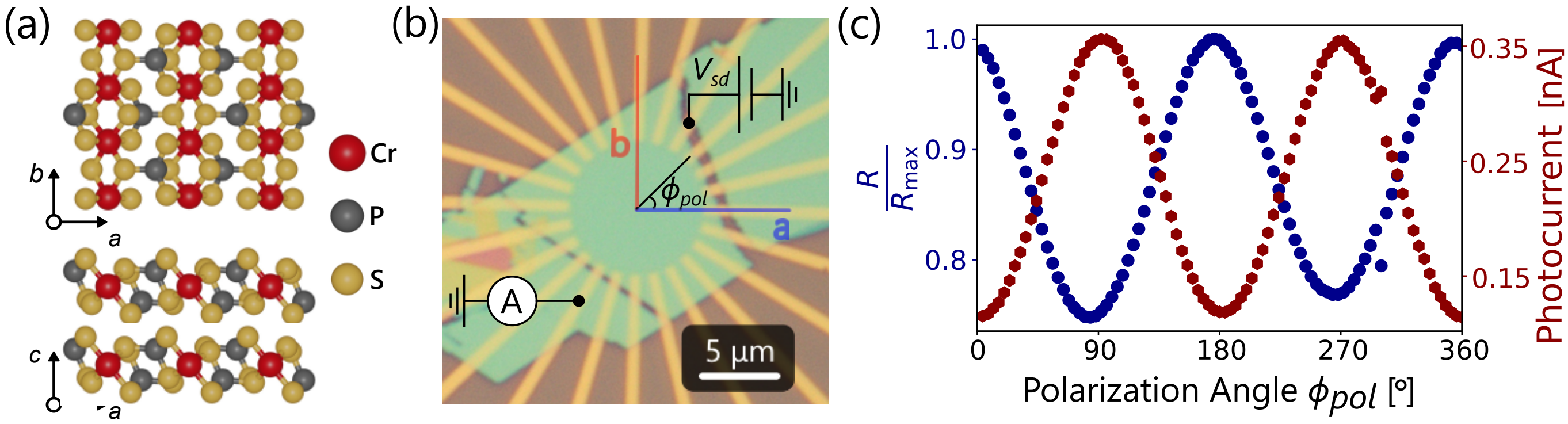

FIG. 1. $CrPS_4$ crystal, optical and optoelectronic anisotropies. (a) Representative crystal structure of a $CrPS_4$ crystal in the *a*-*b* and *a*-*c* planes. (b) Optical micrograph of one of the $CrPS_4$ devices used. The polarization angle of the light with respect to the a-axis of the crystal is labelled as $\phi_{pol}$. The photocurrent measurement configuration is also sketched for an applied source-drain voltage ($V_{\rm sd}$) between two opposite contacts. (c) Normalized reflectivity (in blue) and photocurrent (in red) as a function of $\phi_{pol}$ measured at $E_{\rm ex}$ ≈1.77 eV. The photocurrent measurement was performed at $V_{\rm sd}$= 4 V.

## II. RESULTS AND DISCUSSION

In order to study the optoelectronic properties of the $CrPS_4$, we mechanically exfoliated a bulk crystal onto $Si/SiO_2$ wafers. Using standard lithography and evaporation techniques we fabricated devices with Ti/Au contacts in a circular configuration as shown in Figure 1b.

To precisely determine the crystallographic orientation of our devices, we performed polarized Raman spectroscopy. We assigned the *a*- and *b*- axes according to the polarization-dependent bands (see Figure S1), as previously described and reported in literature.[28,30,34,35] Figure 1b shows the *a*- and *b*- axes for one of the measured devices. The reflectivity of the sample was measured in the middle of the device, where there is minimum influence from the electrodes on the reflection signal. For the photocurrent measurements, we positioned the laser spot close to the area of highest photocurrent, usually in close proximity with the Ti/Au contact-$CrPS_4$ junction, as determined by preliminary photocurrent scans. To increase the signal-to-noise ratio, we measured both the reflectivity and photocurrent using lock-in techniques. The incident light was modulated using an optical chopper, and the reference frequency was used to detect the reflected light at a photodiode. All measurements were performed at room temperature and high vacuum (~1x$10^{-6}$ mbar). The applied source-drain voltages are specified when necessary. Additional details of the device fabrication are provided in the Methods section.

The polarization-dependent measurements of the reflectivity and photocurrent allow us to determine the dichroic response of the $CrPS_4$ device at specific excitation energies ($E_{\rm ex}$). Figure 1c shows the measured reflectivity and photocurrent for an excitation energy of $E_{ex} \approx 1.77$ eV, where a clear 180°–periodicity (consistent with previous polarization dependent microscopy measurements)[28] can be observed both in reflection and photocurrent. At this energy, we observe a maximum in reflection along the a-axis ($\phi_{pol} = 0°$) while for photocurrent the strongest response is along the b-axis ($\phi_{pol} = 90°$). To rule out any contribution to the polarization coming from the $SiO_2$ substrate, we performed a polarization-dependendent spectra for the $CrPS_4$ flake and on the $SiO_2$ substrate (see Figure S2), the latter which displays a negligible contribution to our observations.

The response as a function of the polarization angle, for both reflection and photocurrent, is well described by the function:

$$I\left(\phi_{pol}\right) = A \cdot \cos\left(2\phi_{pol}\right) + I_0, \quad (1)$$

with $\phi_{pol} = \phi - \phi_0$, $\phi$ the incident polarization of the light in our experimental set-up and $\phi_0$ the angle offset with respect to the *a*-axis of the $CrPS_4$. We denote $A$ as the amplitude of the response and $I_0$ as the offset, corresponding to the polarization independent contribution. The linear dichroism is then defined as the difference in reflectivity intensity, or photocurrent signal, measured for the two orthogonal linear polarizations – the *a*-axis, $I_a = I\left(\phi_{pol} = 0°\right)$, and the *b*-axis, $I_b = I\left(\phi_{pol} = 90°\right)$ – divided by their sum:

$$LD = \frac{I_a - I_b}{I_a + I_b} = \frac{A}{I_0}. \quad (2)$$

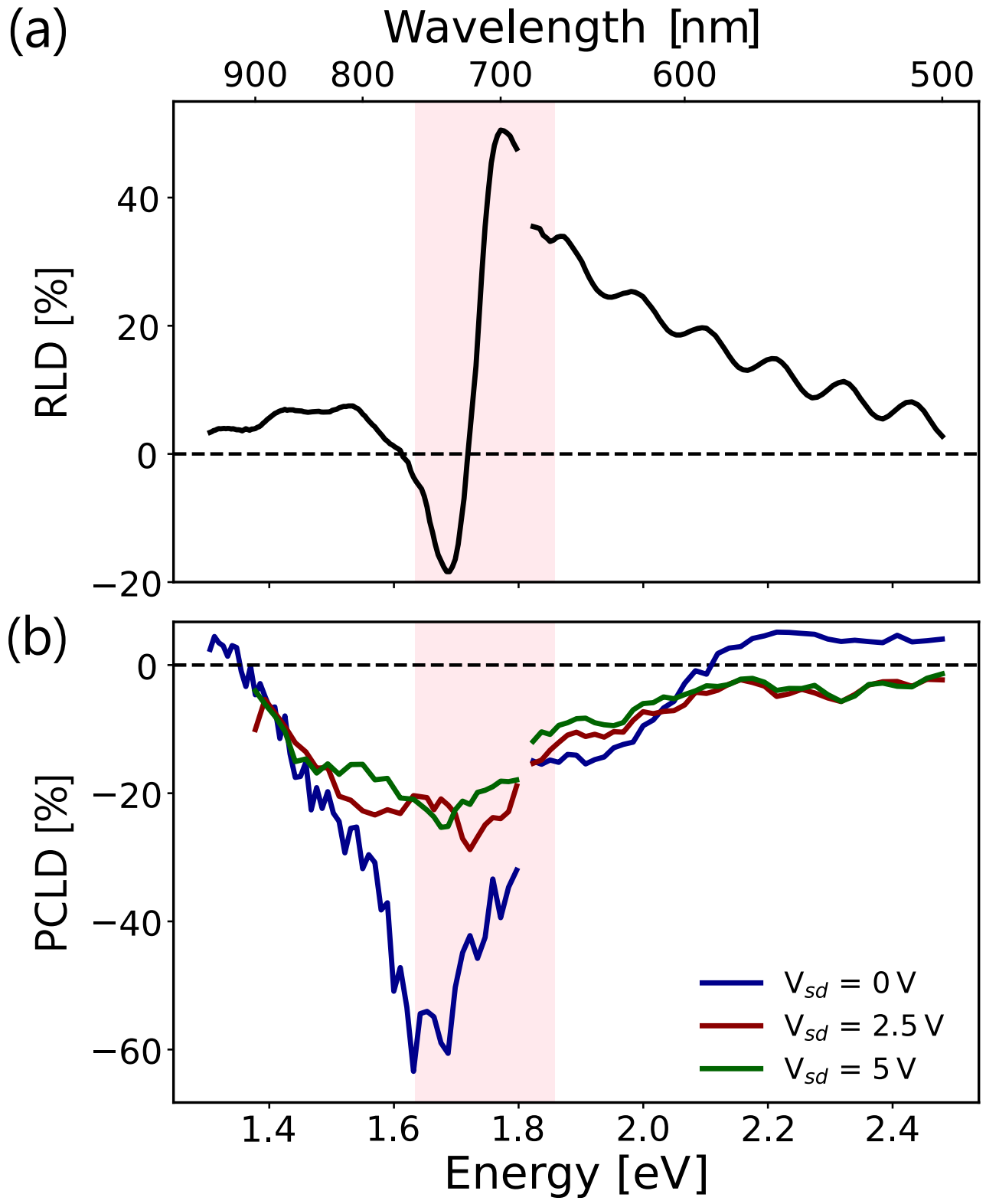

FIG. 2. Reflectivity and photocurrent linear dichroism spectra of the $CrPS_4$ device as calculated from eq. 2. (a) RLD as a function of the wavelength as measured at the center of the device. (b) PCLD as a function of the wavelength for different source-drain voltages. The spectral region with maximum dichroic response for both the RLD and the PCLD is highlighted in both panels.

The effect of the low symmetry in $CrPS_4$ over its optical properties, and how it is related to its crystal structure, can be determined by linear dichroism spectroscopy. To determine this, we have acquired the RLD and PCLD spectra for energies ranging from ∼1.37 eV (950 nm) to ∼2.48 eV (500 nm), shown in Figure 2a and b, respectively. Two distinct features with opposite sign can be observed in the RLD spectra. At lower energies, ∼1.68 eV (738 nm), the RLD reaches ∼ -20% polarization contrast, indicating an increased reflectivity along the *b*-axis as compared to the *a*-axis. For energies around ∼1.77 eV (700 nm), the RLD reaches ∼+50% dichroic response, pointing to a higher reflectivity along the a-axis. This yields a maximum difference of ∼70% in a ∼100 meV range, making it extremely attractive for narrow energy polarization-dependent photodetector applications.

In our measurements, the RLD spectra exhibit their maximum dichroic response in the same energy range as the previously reported absorption spectra of $CrPS_4$. In this spectral region, absorption peaks at 1.6 eV and 1.8 eV (referred to as the $T_1$ and $T_2$ transitions, respectively) arise from $Cr^{3+}$ *d*-*d* optical excitation processes from the $^4A_{2g}$ ground state to the $^4T_{2g}$ and $^4T_{1g}$ excited states respectively.[27–29] The precise location of the $T_1$ and $T_2$ transitions in absorption can shift in energy for a variety of reasons; strain, thickness and/or temperature, for example. Particularly, a change in the band structure with consequences in the optical absorption, is expected as a function of the thickness of the crystal.[28] Despite not observing a clear shift in peak position, we observe a clear enhancement of the dichroic response in thicker devices (see Figure S3 in the Supporting Information).

In order to determine whether or not the RLD features are directly associated with the $T_1$ and $T_2$ transitions, we perform photoluminescence excitation measurements (PLE). From these measurements, the total emitted light can be directly related to the absorption, giving us a clear indication of the nature of our RLD signal.[36,37] Figure S4 in the Supporting information shows the extracted PLE spectra, as well as the resulting degree of polarization. Below 2.2 eV two peaks can be observed at ∼1.7 eV and ∼1.76 eV, consistent with absorption features and our RLD spectra. Above 2 eV, a third transition ($T_3$) has been reported in bulk crystals and assigned to ligand-to-metal charge-transfer, other *d*-*d* transition or a mixture of them.[19,27,38] Our PLE measurements reveal that this feature is polarization independent and therefore not directly visible in our RLD measurements.

While the reflection and PLE spectroscopy give a fingerprint of the material's absorption, photocurrent measurements can carry additional information of the generated charge carriers.[39,40] The quantity of absorbed photons determines the amount of the photogenerated carriers, which can be extracted through the device's contact and quantified as photocurrent.[17] The precise mechanism by which the photocurrent is generated cannot exactly be determined from our measurements. Nonetheless, as most of the photocurrent is located at the interface between the metallic contact and the $CrPS_4$, (see Figure 3b), the photothermoelectric effect (PTE) and the photovoltaic effect (PVE) are the most probable mechanisms at play.[41] A discussion on the possible mechanisms is presented in the Supporting Information section .

A strong dependence of the PCLD with the excitation energy is observed at different source-drain voltage ($V_{sd}$) values (Figure 2b), reaching ∼ 60% at $E_{ex} \approx 1.63$ eV at $V_{sd}$ = 0V. As the photocurrent intensity is proportional to the amount of photogenerated electron-hole pairs, an increased photoresponse is expected close to the absorption band. This feature has also been observed in direct absorption[28] and unpolarized photocurrent measurements[32] in samples from 20 nm thickness down to a single layer. From our measurements, we observe a clear polarization preference for this transition along the *b*-axis, where the magnitude of the photocurrent is significantly increased.

When applying a $V_{sd}$ bias we observe a broadening in the peak of the PCLD. As the dark current increases together with the $V_{sd}$, a bias-dependent background lowers the relative PCLD magnitude to ∼ 25%. As the applied $V_{sd}$, is increased, the responsivity of the device rises as well, nonetheless, the amplitude and offset increase equivalently, such that the overall resulting PCLD spectra remains unchanged. Figure S7 shows the photocurrent as a function of $\phi_{pol}$ for a few $V_{\mathrm{sd}}$ values. The $A$ and $I_0$ parameters, as well as the calculated PCLD values, are extracted for each plot and summarized in Table S1, in the Supporting Information.

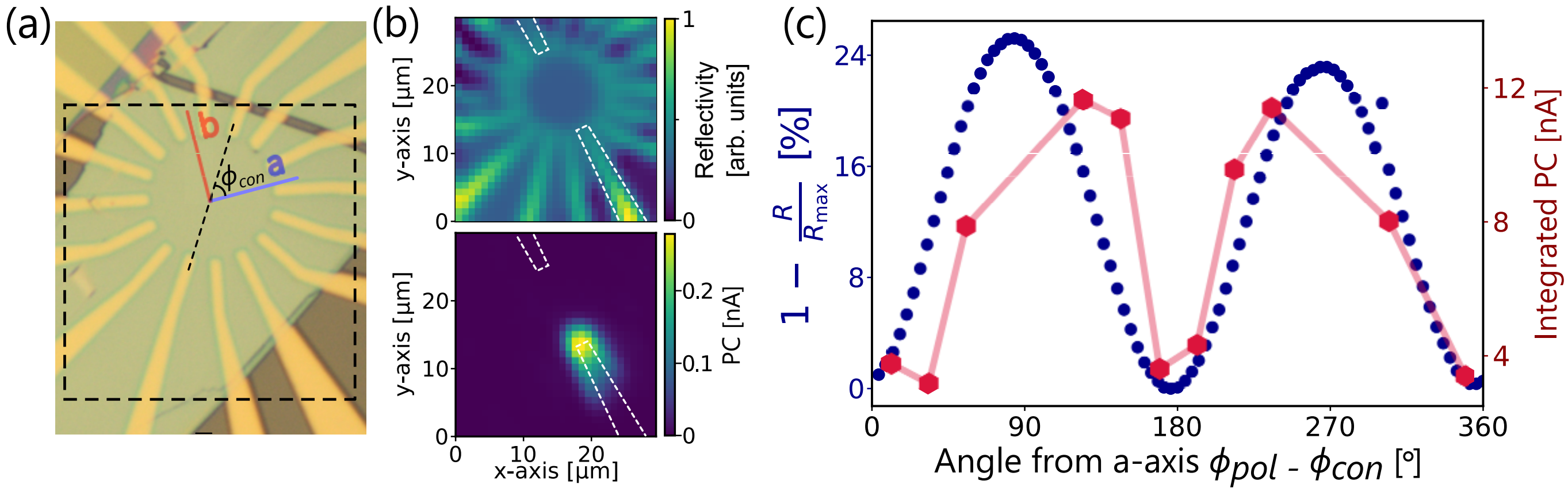

FIG. 3. Photocurrent intensity along different crystallographic directions. a) Optical micrograph of device used for the measurements. The axes of the crystal are indicated in blue and red for the $a$- and $b$- axes, respectively. The dashed square is the approximate area where the photocurrent scans were measured. b) Reflectivity (top) and photocurrent (bottom) scans for one set of opposite contacts measured at $V_{\mathrm{sd}}$= 4 V. c) Integrated photocurrent for a set of 11 circular contacts as extracted from the photocurrent map from panel b). The absorption spectra (plotted as 1-$R/R_{\mathrm{max}}$) is also shown to provide a general trend comparison. Both the reflectivity and photocurrent measurements are performed at excitation energy $E_{\mathrm{ex}} \approx 1.77$ eV.

The most significant difference between the RLD and the PCLD spectra is the lack of change of sign in the photocurrent spectra. We attribute this to the intrinsic nature of photocurrent, where firstly electron-hole pairs are created, dissociated by an electric field and collected at the contacts as a photocurrent.[17] The photocurrent difference for orthogonal polarizations captures the anisotropy in absorption, which is proportional to the extinction coefficient; whilst the reflection is related to the complex refractive index.[42] Our LD measurements suggest that anisotropy of the complex refractive index changes sign as a function of the excitation energy, whilst the extinction coefficient maintains its polarity.

To discern how the crystal anisotropy affects the optoelectronic response in our device, we perform photocurrent maps along polar opposite contacts of our device, as displayed in the electrical diagram in Figure 3a. By changing the contact pair, *i.e.* the angle $\phi_{con}$, within the a-b plane in the crystal, we selectively probe the photoresponse along different crystallographic directions. Figure 3b shows the reflectivity and photocurrent maps with the contact pair along the b-axis with $V_{\mathrm{sd}}$= 4 V. In both panels, the position of the contacts is outlined by a white dashed line for clarity. From the photocurrent maps we extract the integrated photocurrent (total photocurrent generated) for a set of 11 contacts along different crystallographic angles as shown in Figure 3c. We obtain the lowest integrated photocurrent for the contacts aligned with the $a$-axis and the highest for the contacts along the $b$-axis. The shift and overall shape variations in comparison to the absorption (plotted as 1-$R/R_{max}$) can be a result of the local differences at each contact-$CrPS_4$ junction, the strain between the flake and the contact, amongst others. This anisotropic photoconductive property has been observed in other low-symmetry vdWs materials, in different spectral ranges,[43,44] nonetheless, the strong optical anisotropy and the functionality at room-temperature, makes of $CrPS_4$ a strong candidate for integration in more complex heterostructures for a wide range of applications.

## III. CONCLUSIONS

Our measurements demonstrate the strong optical anisotropy in $CrPS_4$ and how it is linked to its crystal axes. The strong linear polarization modulation between 1.6 eV and 1.9 eV, together with the change in sign of the RLD at room-temperature, can be exploited for narrowband photodetector applications at room-temperature. Additionally, our scanning photocurrent measurements reveal a clear modulation in the photoresponse of the $CrPS_4$ device along different crystallographic directions of the material. As has been recently shown in $CrPS_4$-TMDs heterostructures, materials with different symmetries can be used to induce nonlinear photocurrents and layer-dependent control over valley polarization in proximitized heterostructures.[33,45] Additionally, ultrafast optical switches have been proposed using $CrPS_4$-based devices.[31] Our measurements demonstrate the relevance of linear polarization spectroscopy and the enhanced photoresponse along the $b$-axis in $CrPS_4$, showcasing the strong optoelectronic anisotropy at room temperature. We envision that these functionalities can be implemented in future 2D spintronic devices, coupling to the magnetic lattice for magnetisation dynamics and proximitized applications with other 2D vdWs materials.

## IV. METHODS

### Device Fabrication

The $CrPS_4$ flakes are obtained by mechanical exfoliation (bulk crystal supplied by HQ Graphene) on a $Si/SiO_2$ (285 nm) substrate in a nitrogen environment. Using an optical microscope, the $CrPS_4$ flakes are selected based on their size, thickness, and homogeneous surface. Using standard lithography techniques, the Ti/Au (5 nm/45 nm) contacts are fabricated on top of the flake by means of electron beam lithography and electron beam evaporation. The thicknesses of the devices is characterized using a Multimode Atomic Force Microscope from Bruker.

### Optoelectronic measurements

To determine the crystallographic axes, raman spectra are obtained with an inVia Raman Renishaw microscope using a linearly polarized laser in backscattering geometry. The excitation wavelength and grating used are $\lambda$ = 532 nm and 1800 l/mm respectively. The laser power was kept below ~100 µW with a diffraction-limited spot of ~1 µm positioned in the middle of the finished device. The polarization dependence was obtained by rotating the sample in intervals of ~10° for each measurement.

For the photocurrent, reflectivity and photoluminescence excitation measurements, a supercontinuum white light laser (NKT Photonics SuperK EXTREME) is used as the illumination source. The laser light is polarized and focused using a 10x achromatic objective. The samples were mounted in a Janis ST-500 flow cryostat at high vacuum (~1x10$^{-6}$ mbar). All the measurements were performed at room temperature and zero electrostatic gating. The induced photocurrent is measured in a short-circuit configuration using a Stanford Research Systems SR830 lock-in amplifier, which is referenced to the frequency of the optical chopper. The photocurrents are converted to a voltage using a home-build current preamplifier, which is subsequently measured by the lock-in amplifier. The photoluminescence spectra were collected in backscattering geometry through the same objective, filtered with a 800 nm long-pass filter to remove residual excitation light, and analyzed using an Andor Shamrock 500i spectrometer, with an iDus 420 thermoelectrically cooled CCD detector with a 300 l/mm grating. The spectra were later normalized by the laser power at each excitation energy.

For all our measurements, $\phi_{pol}$ describes the angle between the linear polarization axis of the incoming light and the a-axis of the $CrPS_4$. The polarization is controlled by rotating a half-waveplate, keeping the sample orientation fixed.

## ACKNOWLEDGEMENTS

C.A.C-S. acknowledges Prof. M. A. Loi, J. Pinna and M. Kot for allowing access and technical support with the Raman microscope. C.A.C-S. acknowledges K. Sundararajan for discussion regarding the $CrPS_4$ band structure and orbital symmetries. C.A.C-S. acknowledges K. Nakagawa for additional discussion on the manuscript. The authors acknowledge as J. G. Holstein, H. Adema, H. de Vries, F. H. van der Velde and A. Joshua, for their technical support. Sample fabrication was performed using NanoLabNL facilities.

## AUTHOR CONTRIBUTIONS

C.A.C-S. and H.M. performed the sample fabrication, electrical, optical measurements and the data analysis under the supervision of M.H.D.G. and help from D.V.. Photoluminescence excitation measurements were performed and analyzed by T.L.C.. C.A.C-S. wrote the paper with the support of D.V. with comments from all the authors under the supervision of M.H.D.G..

## FUNDING SOURCES

This work was supported by the “Materials for the Quantum Age” (QuMat) program (Registration No. 024.005.006) which is part of the Gravitation program financed by the Dutch Ministry of Education, Culture and Science (OCW), the European Union (ERC, 2D-OPTOSPIN, 101076932) and the Zernike Institute for Advanced Materials.

# Supporting Information: Room Temperature Polarization-Resolved Optical and Photocurrent Spectroscopy in Low-Symmetry van der Waals Semiconductor $CrPS_4$

Cédric A. Cordero-Silis[1], Daniel Vaquero[1], Teresa López-Carrasco[1], Harshan Madeshwaran[1], and Marcos H. D. Guimarães[1*]

[1]*Zernike Institute for Advanced Materials, University of Groningen, 9747 AG Groningen, The Netherlands*

E-mail: m.h.guimaraes@rug.nl

# Polarized Raman Spectroscopy

Polarized Raman spectroscopy can be used to determine the crystallographic axes in certain materials. For $CrPS_4$ , the peaks at 168 cm$^{-1}$ and 256 cm$^{-1}$ can be used to identify the *a*- and *b*- axes.[S1] For an excitation wavelength of 532 nm, the b-axis can be identified through the analysis of the 168 cm$^{-1}$ mode Raman intensity as a function of the polarization as shown in Figure S1.

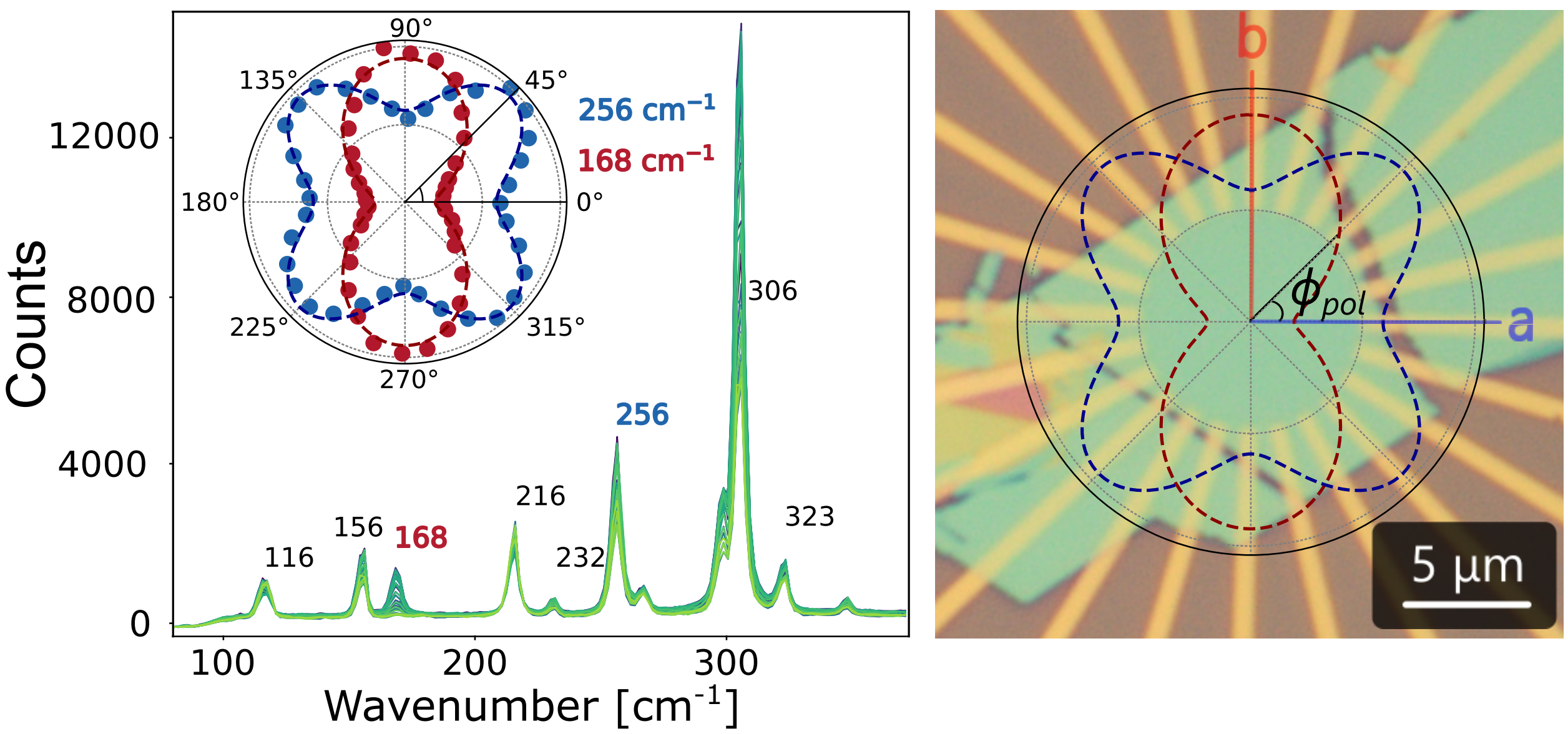

Figure S1: Polarization-resolved Raman spectra of a ∼21 nm thick $CrPS_4$ flake. The inset shows a polar plot with the 168 cm$^{-1}$ and 256 cm$^{-1}$ modes as a function of the polarization angle $\phi_{pol}$ used to assign the crystallographic axes. The microscope image shows the assignment for the *a*- and *b*-axes according to the $2\phi$ and $4\phi$ fit in one of the devices used.

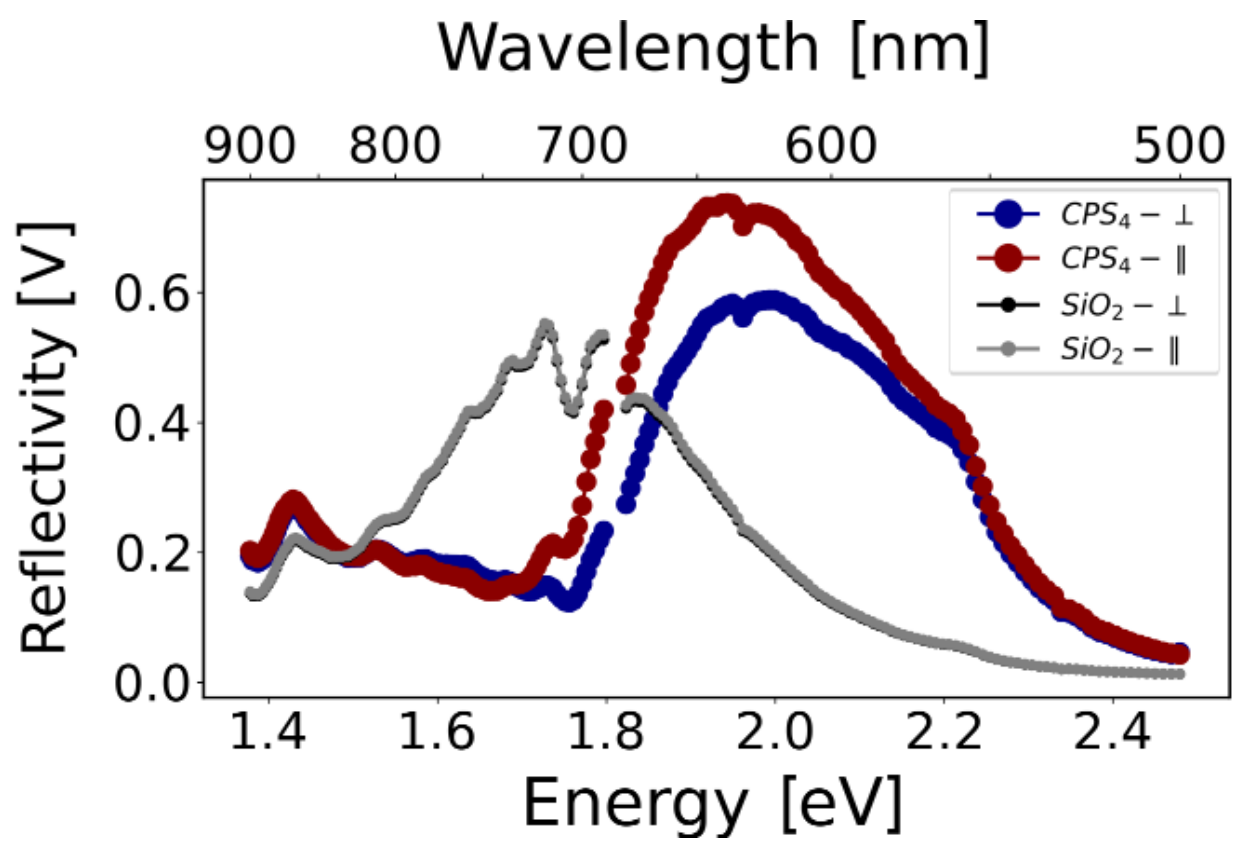

Figure S2: Reflectivity signal of the $CrPS_4$ device and $SiO_2$ substrate with the laser polarizations at $\phi_{pol} = 0°$ ($\parallel$ to the a-axis) and at $\phi_{pol} = 90°$ ($\perp$ to the a-axis). The $SiO_2$ substrate's signal shows minimum polarization dependence.

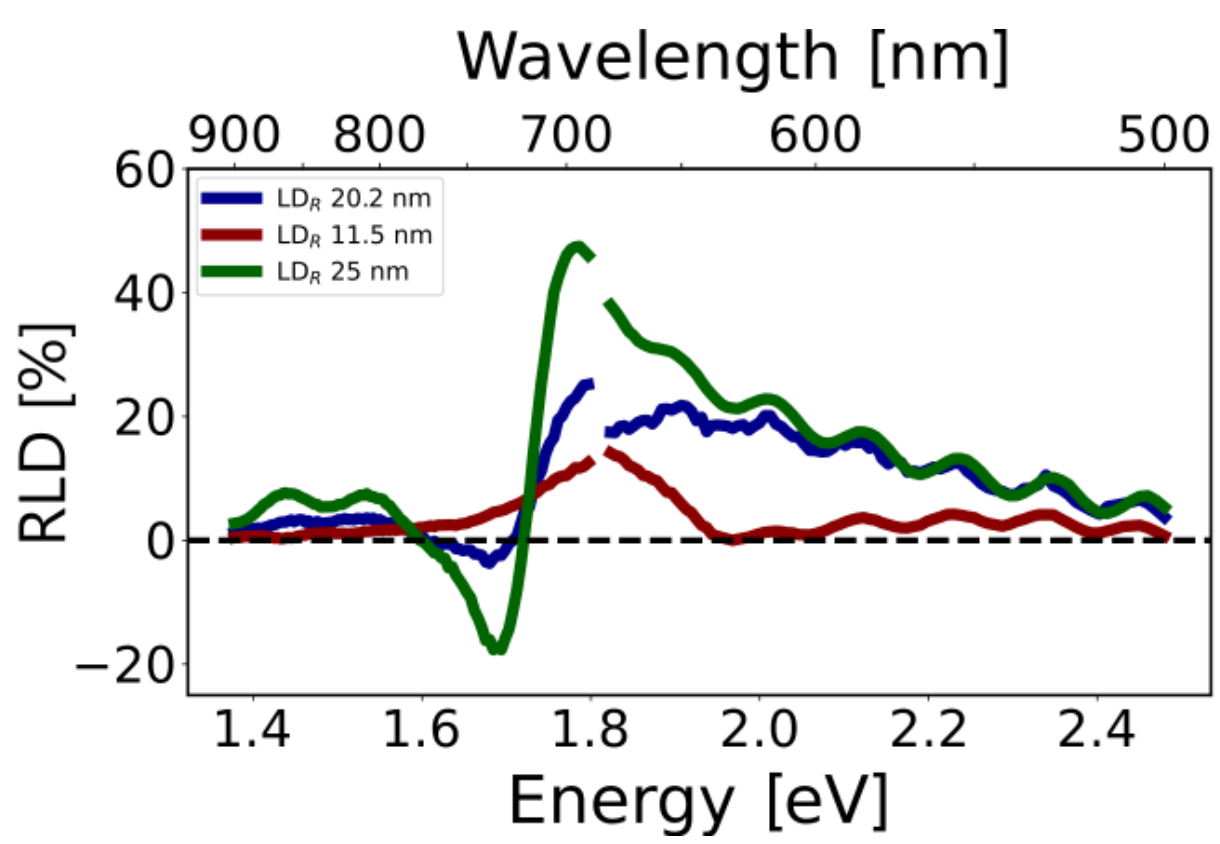

Figure S3: RLD spectra for different sample thicknesses calculated following equation 2. The spectra for the 25 nm thickness correspond to the device in the main text.

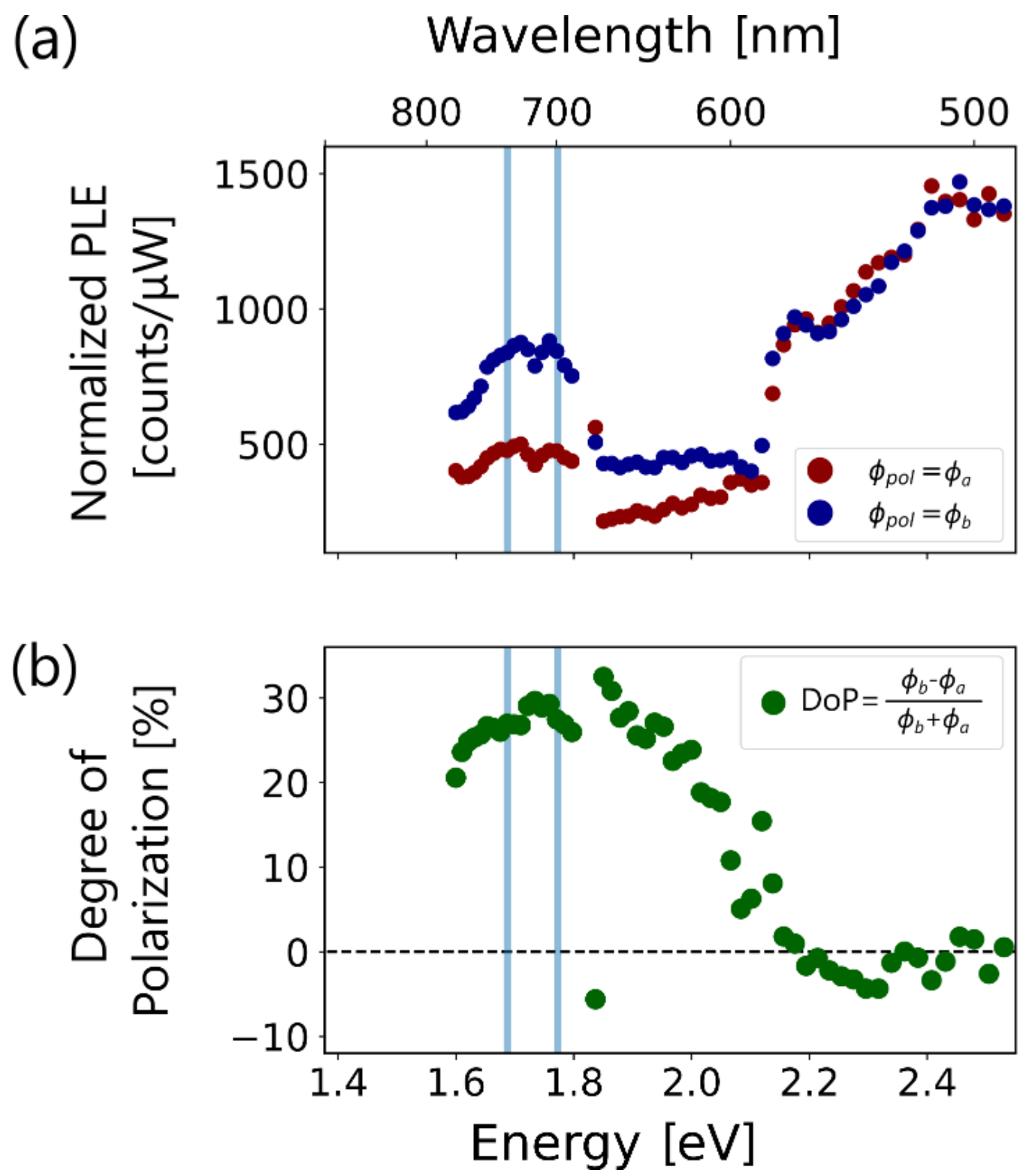

Figure S4: Photoluminescence excitation (PLE) measurements. (a) Integrated PLE area (normalized by laser power) as a function of the wavelength for polarization along the $a$-axis ($\phi_{\rm pol} = \phi_{\rm a}$) and along the $b$-axis ($\phi_{\rm pol} = \phi_{\rm b}$). (b) Degree of polarization as function of the excitation wavelength. In both panels, the energy at which the RLD spectra shows a dip or a peak (see Figure 2 in the main text) is highlighted, ∼1.68 eV and ∼1.77 eV.

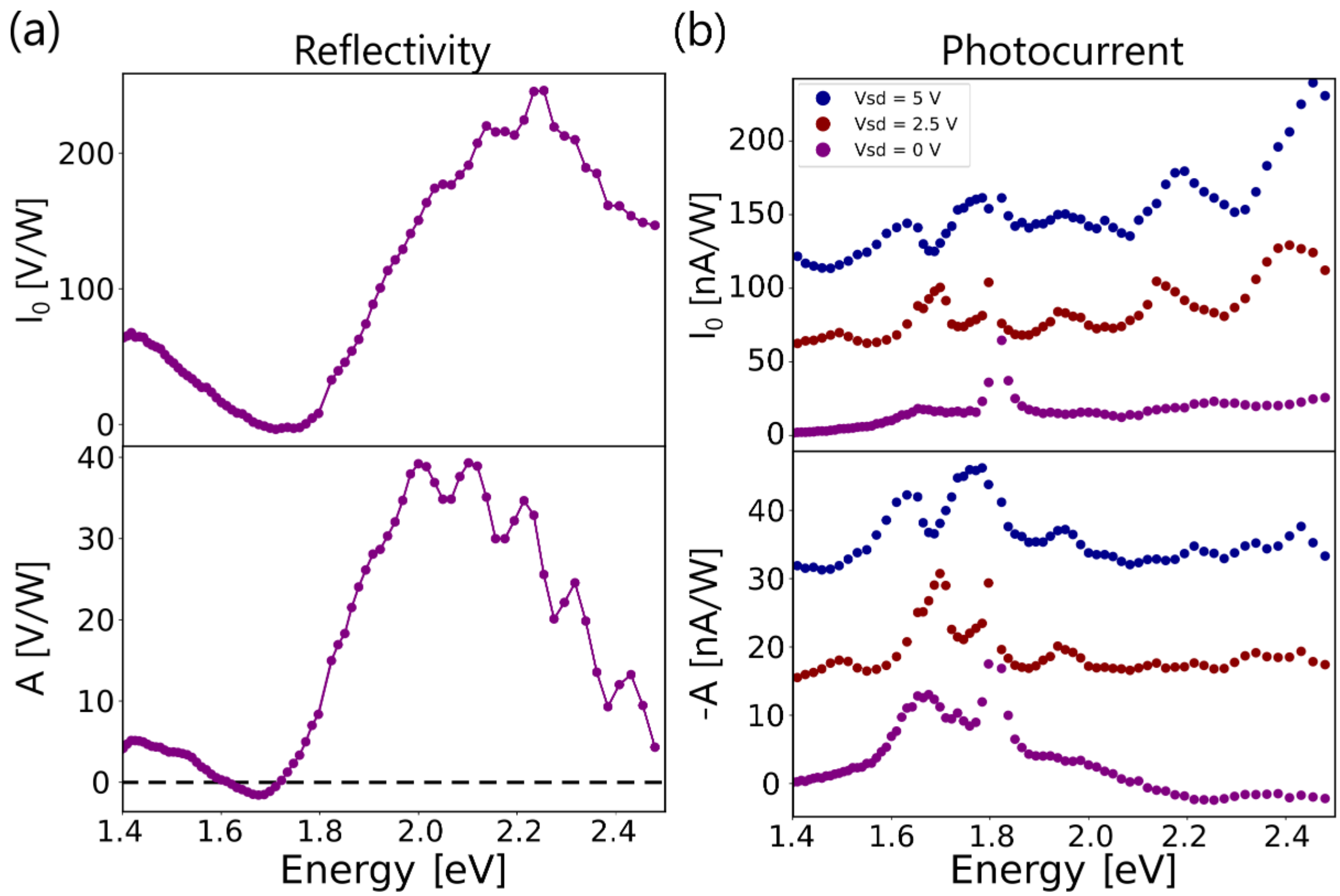

Figure S5: Polarization independent and amplitude from the (a) RLD and (b) PCLD measurements. The photocurrent spectra for the $A$ and $I_0$ parameters, for different $V_{sd}$, values, are offset for clarity.

From fitting equation 1 to the polarization dependence reflection and photocurrent we can extract the polarization independent and amplitude components of both responses. From the reflectivity, panel (a) in Figure S5, a dip in the signal is associated to increased absorption, therefore the decrease in both $I_0$ and $A$ at ∼1.7 eV and after 2.1 eV, can be associated to absorption processes. The parameters extracted from our polarized photocurrent measurements (b) are not as straightforward to interpret. As the laser spot was placed in close proximity to the Ti/Au contact-$CrPS_4$ interface, we might have some additional signal from the metallic contact itself. Nevertheless we observe an increase in the photocurrent amplitude (along the $b$-axis of the $CrPS_4$ device) between 1.6 and 1.8 eV as well as increased $I_0$ signal above 2 eV for $V_{sd} > 0$.

## Photocurrent generation mechanism

In the PTE, a voltage difference ($V_{PTE}$) is generated as a result of the difference in Seebeck coefficient between the two materials in contact, driving a temperature gradient ($\Delta T$) between them. This results in charge carriers flowing from the hot region towards the cold region and can be expressed as $V_{PTE} = (S_2 - S1)\,\Delta T$, where $S_2$ and $S_1$ are the Seebeck coefficients of, respectively, the metal electrode and the material.[S2–S4] Although the Seebeck coefficient for $CrPS_4$ has not yet been reported, we can estimate it from our room temperature measurements. Therefore, considering a $S_2 = 10\frac{\mu V}{K}$ (common for metallic contacts) and $S_1 \approx 350 - 800\frac{\mu V}{K}$ (estimation for 2D semiconductors), and assuming 50% of the photocurrent originating due to the PTE, we obtain a $\Delta T$ of $\sim$ 2530 K to 5880 K.

On the other hand, in the photovoltaic effect (PVE), a large built-in electric field, generated by the presence of a Schottky barrier at the metal-insulating interface, induces the separation of the photogenerated electron-hole pairs, resulting in a photocurrent.[S5,S6] Additionally, the intensity of the electric field within the depletion region in a metal-semiconductor interface is larger within the semiconductor area. As a result of this, the maximum photocurrent signal can shift from the contact-flake interface towards the $CrPS_4$ when increasing the source-drain voltage. From our scanning photocurrent measurements, we observe an increased PC signal within the semiconductor area that aligns with this mechanism (see Figure S6).[S7] However, in order to determine the photocurrent generation mechanism with more certainty, power-dependent measurements, as well as transport measurements (via local gating, for example), need to be performed. These can help confirm or distinguish it from other mechanisms, as well as quantify the contribution from each effect to the total photocurrent.[S4,S8,S9]

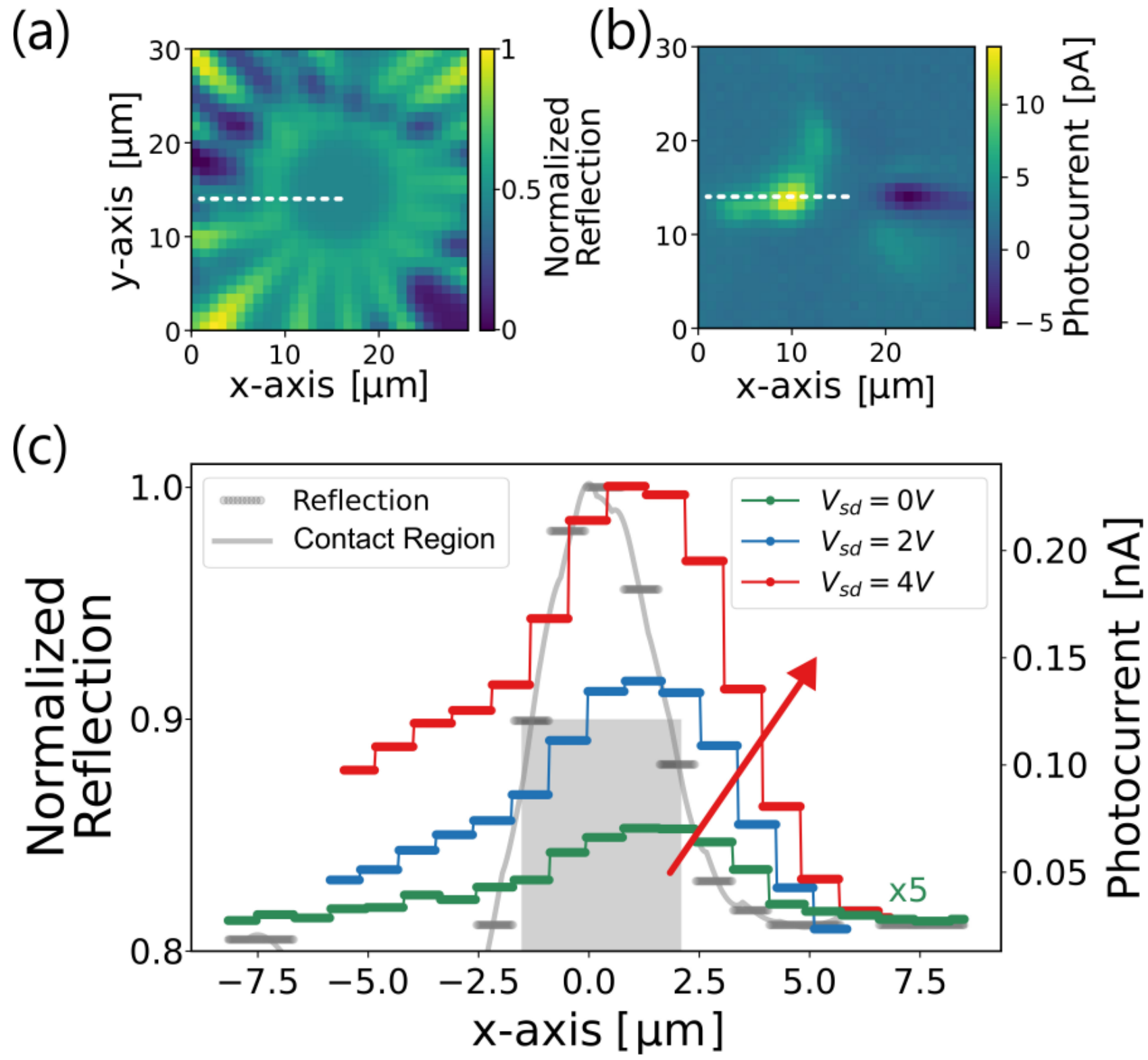

Figure S6: Photocurrent location and intensity as a function of the source-drain bias. (a) Reflectivity and (b) photocurrent maps of the device at $V_{sd} = 0V$. The dashed white line is the line trace at which the photocurrent traces are shown in panel (c). The contact position is represented by a grey rectangle. A slight shift and an increase in PC, towards the centre of the device, is schematically shown by the coloured arrow. Note that the PC is increased by 5 for the $V_{sd} = 0$ measurement.

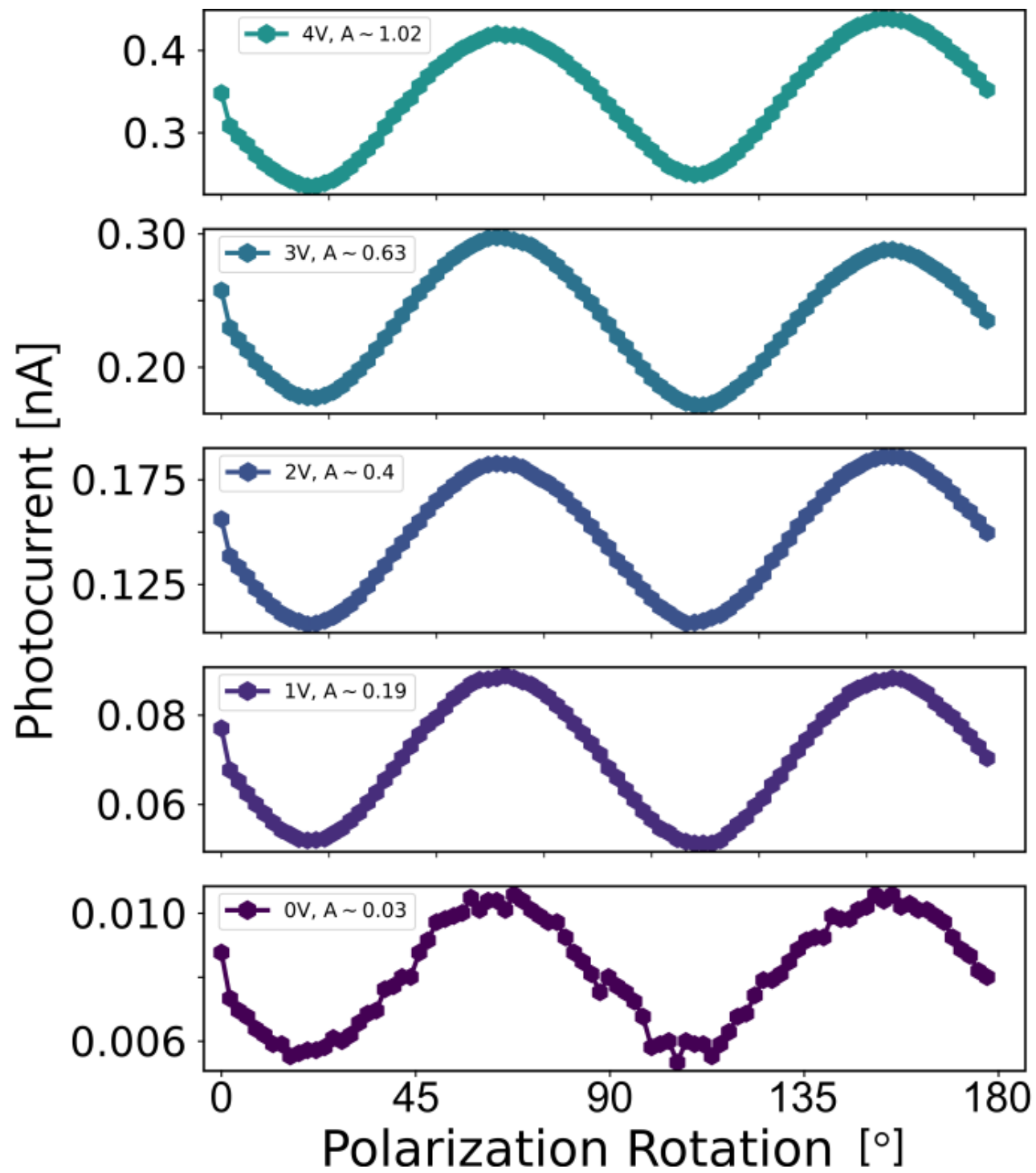

Figure S7: Photocurrent as a function of the polarization at different $V_{sd}$ values with $E_{ex} \approx 1.82$ eV. The measurements are taken from the device shown in Fig. 3 panel (a). The approximate values of the amplitude, offset and their associated linear dichroism are shown in Table S1.

Table S1: Extracted parameters (from equation 1) for each polarization dependence at each $V_{sd}$ value in Figure S7.

| $V_{sd}$ [V] | A [nA] | $I_0$ [nA] | LD [%] |
|---|---|---|---|
| 0 | 0.03 | 0.01 | 13.0 |
| 1 | 0.19 | 0.07 | 11.9 |
| 2 | 0.40 | 0.15 | 11.7 |
| 3 | 0.63 | 0.24 | 11.6 |
| 4 | 1.02 | 0.34 | 13.0 |